\documentclass[12pt]{article} 
\usepackage{bbold}
\usepackage{cite}
\usepackage{graphicx}
\def \bea{\begin{eqnarray}} 
\def \beq{\begin{equation}}

\def \eea{\end{eqnarray}} 
\def \eeq{\end{equation}}

\def \half{\frac{1}{2}}
\def \s{\sqrt{2}} 
\def \st{\sqrt{3}} 
\def \sx{\sqrt{6}} 
\def\lsim{\mathrel{\rlap{\lower3pt\hbox{$\sim$}}\raise2pt\hbox{$<$}}}
\def\gsim{\mathrel{\rlap{\lower3pt\hbox{$\sim$}}\raise2pt\hbox{$>$}}}

\begin{document} 
\begin{flushright}
TECHNION-PH-2016-05 \\
EFI 16-09 \\
March 2016 \\
\end{flushright} 
\centerline{\bf From $\Xi_b \to \Lambda_b \pi$ to $\Xi_c \to \Lambda_c \pi$}
\medskip
\centerline{Michael Gronau}
\centerline{\it Physics Department, Technion, Haifa 32000, Israel}
\medskip 
\centerline{Jonathan L. Rosner} 
\centerline{\it Enrico Fermi Institute and Department of Physics,
  University of Chicago} 
\centerline{\it Chicago, IL 60637, U.S.A.} 
\bigskip

\begin{quote}
Using a successful framework for describing S-wave hadronic decays of light
hyperons induced by a subprocess $s \to u (\bar u d)$, we presented recently
a model-independent calculation of the amplitude and branching ratio for
$\Xi^-_b \to \Lambda_b \pi^-$ in agreement with a LHCb measurement. The same
quark process contributes to $\Xi^0_c \to \Lambda_c \pi^-$, while a second term
from the subprocess $cs \to cd$ has been related by Voloshin to differences
among total decay rates of charmed baryons. We calculate this term and find it
to have a magnitude approximately equal to the $s \to u (\bar u d)$ term.
We argue for a negligible relative phase between these two contributions, 
potentially due to final state interactions.  However, we do not know whether
they interfere destructively or constructively.  
For constructive interference
one predicts ${\cal B}(\Xi_c^0 \to \Lambda_c \pi^-) = (1.94 \pm 0.70)\times 10^{-3}$
and ${\cal B}(\Xi_c^+ \to \Lambda_c \pi^0) = (3.86 \pm 1.35)\times 10^{-3}$.  
For destructive interference, the respective branching fractions are
expected to be less than about $10^{-4}$ and $2 \times 10^{-4}$.
\end{quote}

\leftline{\qquad PACS codes: 14.20.Lq, 14.20.Jn, 12.40.Nn, 13.30.Eg}
\bigskip

\section{INTRODUCTION} \label{sec:intro}

Most decays of charmed and beauty baryons observed up to now occur 
by $c$ and $b$ quark decays. In strange heavy flavor baryons an $s$ quark 
may decay instead via the heavy flavor conserving subprocess $s \to u (\bar u d)$ 
or $su \to ud$, with the $c$ or $b$ quark acting as a spectator. In strange charmed baryons 
an additional Cabibbo-suppressed subprocess $c s \to c d$ can contribute. Early investigations 
of heavy flavor conserving two body hadronic decays of charmed and beauty baryons involving 
a low energy pion have been performed in
Ref.~\cite{Cheng:1992ff,Sinha:1999tc,Voloshin:2000et,Li:2014ada,Faller:2015oma,Cheng:2015ckx}.  
In these studies a soft pion limit, partial conservation of the axial-vector current (PCAC) and current algebra have implied expressions for decay amplitudes in terms of matrix elements of four-fermion operators between initial and heavy baryon states. 
These matrix elements are difficult to estimate and depend strongly on models for heavy baryon wave functions.  

Recently we proposed a model-independent approach for studying the decay 
$\Xi_b^- \to \Lambda_b \pi^-$~\cite{Gronau:2015jgh} which had just been observed 
by the LHCb collaboration at CERN~\cite{Aaij:2015yoy}. In the heavy $b$ quark limit 
this decay by $s \to u (\bar u d)$ proceeds purely via an S-wave. Assuming  that properties 
of the light diquark in $\Xi_b^-$ are not greatly affected by the heavy nature of the spectator 
$b$ quark, the decay amplitude for $\Xi_b^- \to \Lambda_b \pi^-$ may be related to amplitudes 
for S-wave nonleptonic decays of $\Lambda$, $\Sigma$, and $\Xi$ which have been measured 
with high precision~\cite{Roos:1982sd}.  We  calculated a branching fraction for 
$\Xi^-_b \to \Lambda_b \pi^-$ consistent with the range 
allowed in the LHCb analysis. Our purpose now is to extend this calculation to charmed baryon 
decays $\Xi_c^0 \to \Lambda_c \pi^-$ and $\Xi_c^+ \to \Lambda_c \pi^0$.  

Sec. \ref{sec:supi} summarizes the result of Ref.~\cite{Gronau:2015jgh} for the 
amplitude of $\Xi^-_b \to \Lambda_b \pi^-$, in which the underlying quark transition is
$s \to u (\bar u d)$. This result is then applied to a contribution of the same quark 
subprocess to $\Xi^0_c \to \Lambda_c \pi^-$. 
A second term in this amplitude due to the subprocess $c s \to c d$
is studied in Sec.~\ref{sec:cscd}. The total amplitude and the branching ratios for 
$\Xi^0_c \to \Lambda_c \pi^-$ and $\Xi_c^+ \to \Lambda_c \pi^0$ are calculated in 
Sec. \ref{sec:total} while Section \ref{sec:con} concludes.

\section{$s \to u (\bar u d)$ TERM IN $\Xi^-_b \to \Lambda_b \pi^-$ AND
$\Xi^0_c \to \Lambda_c \pi^-$}
\label{sec:supi}

We will use notations which are common for describing hadronic hyperon 
decays~\cite{Roos:1982sd}. The effective Lagrangian for $B_1 \to B_2 \pi$ given by
\beq\label{eqn:SP}
{\cal L}_{\rm eff} = G_F m_\pi^2[\bar \psi_2(A + B \gamma_5)\psi_1] \phi_\pi
\eeq
involves two dimensionless parameters $A$ and $B$ describing S-wave and P-wave
amplitudes, respectively. Here $G_F = 1.16638 \times 10^{-5}$ GeV$^{-2}$ is the Fermi 
decay constant. The partial width is
\beq \label{eqn:rate}
\Gamma(B_1 \to B_2 \pi ) = \frac{(G_F m_\pi^2)^2}{8 \pi m_1^2} q
 [(m_1+m_2)^2-m_\pi^2]|A|^2 + [(m_1-m_2)^2-m_\pi^2]|B|^2~,
\eeq
where $q$ is the magnitude of the final three-momentum of either particle in the
$B_1$ rest frame.

Consider first $\Xi_b^- \to \Lambda_b \pi^-$ studied in Ref. \cite{Gronau:2015jgh}.
In the heavy $b$ quark limit the light 
quarks $s$ and $d$ in $\Xi^-_b=bsd$ are in an S-wave state anstisymmetric in flavor 
with total spin $S = 0$. The light quarks $u$ and $d$ in the $\Lambda_b = bud$ are 
also in an S-wave
state with $I = S = 0$.  In the decay $\Xi^-_b \to \Lambda_b \pi^-$ which proceeds via 
$s \to u (\bar u d)$ the $b$ quark acts as a spectator.
The transition among light quarks is thus one with $J^P = 0^+ \to 0^+ \pi$,
and hence is purely a parity-violating S wave.  Thus it may be related to parity-violating 
S-wave amplitudes in nonleptonic decays of the hyperons $\Lambda$, $\Sigma$, and $\Xi$.

S-wave hadronic decays of hyperons, $B_1 \to B_2 \pi$, where the baryons $B_1$ and $B_2$ 
belong to the lowest SU(3) octet baryons, have been known for fifty years to be described well 
by using PCAC and current algebra and assuming octet dominance 
\cite{Sugawara:1965zza,Suzuki:1965zz}.  An equivalent and somewhat more compact 
parametrization of these amplitudes based on duality has been suggested a few years 
later~\cite{Nussinov:1969hp}. All hyperon S-wave amplitudes may be expressed in terms of an overall normalization parameter $x_0$ and a parameter $F$ describing the ratio of antisymmetric and symmetric three-octet coupling. (In the soft pion limit the commutator of the axial charge with the weak Hamiltonian represents a 
third octet in addition to the two baryons.) Thus one 
finds~\cite{Gronau:2015jgh,Nussinov:1969hp}
\bea\label{eqn:amps}
A(\Lambda \to p \pi^-) & = & -(2F+1) x_0/\sx~,\nonumber \\
A(\Sigma^+ \to n \pi^+) & = & 0~,\nonumber \\
A(\Sigma^- \to n \pi^-) & = & -(2 F - 1) x_0~,\nonumber\\
A(\Xi^- \to \Lambda \pi^-) & = & (4F -1) x_0/\sx~,
\eea
while amplitudes involving a neutral pion are related to these amplitudes by
isospin.  Using best fit values $F = 1.652,~x_0= 0.861$, one finds good agreement
between predicted and measured amplitudes as shown in Table I (see
\cite{Gronau:2015jgh}).  The relative signs of S-wave amplitudes are
convention-dependent and differ from those in Ref.\ \cite{Roos:1982sd}.  An
overall sign change is also permitted, associated with two possible signs of $x_0$.

\begin{table}
\begin{center}
\caption{Predicted and observed S-wave amplitudes $A$ for nonleptonic hyperon
decays.  Predicted values are for best-fit parameters $F = 1.652$, $x_0 =
0.861$.
\label{tab:amps}}
\begin{tabular}{c c c c} \hline \hline
     Decay    & Predicted $A$ & Observed & Predicted \\
              &   amplitude   &  value~\cite{Roos:1982sd}  &   value   \\ \hline
$\Lambda \to p \pi^-$ & $-(2F+1)x_0/\sx$ & $-1.47 \pm 0.01$ & --1.51 \\
$\Lambda \to n \pi^0$ & $(2F+1)x_0/(2\st)$ & $ 1.07 \pm 0.01$ & 1.07 \\
$\Sigma^+ \to n \pi^+$ &       0      & $0.06 \pm 0.01$ &    0    \\
$\Sigma^+ \to p \pi^0$ & $-(2F-1)x_0/\s$ & $-1.48 \pm 0.05$ & --1.40   \\
$\Sigma^- \to n \pi^-$ &  $-(2F-1)x_0$   & $-1.93 \pm 0.01$ & --1.98   \\
$\Xi^0 \to \Lambda \pi^0$ & $(4F-1)x_0/(2\st)$ & $1.55 \pm 0.03$ & 1.39 \\
$\Xi^- \to \Lambda \pi^-$ & $(4F-1)x_0/\sx$ & $2.04 \pm 0.01$ & 1.97 \\
\hline \hline
\end{tabular}
\end{center}
\end{table} 

In the decay $\Xi^-_b \to \Lambda_b \pi^-$, which also proceeds by $s \to u (\bar u d)$,
the light diquarks $sd$ and $ud$ in the initial and final baryons form each a spinless 
antisymmetric $3^*$ of flavor SU(3). The weak transition occurs between this pair of 
diquarks while the $b$ quark acts as a spectator. Neglecting the effect of the heavy $b$ 
quark on relevant properties of the light diquarks, this amplitude is expected to be equal to 
an amplitude for a transition between light hyperons, $\Lambda \to \Lambda (\bar u u)$, in 
which the diquarks in initial and final hyperons are also in an antisymmetric $3^*$ while the 
$s$ quark acts as a spectator.  Thus one finds~\cite{Gronau:2015jgh}
\beq \label{eqn:xibamp}
A(\Xi^-_b \to \Lambda_b \pi^-) = (5F - 2)x_0/3~.
\eeq
Using the best fit values of $x_0$ and $F$ one obtains $A(\Xi^-_b \to \pi^- \Lambda_b)
= \pm 1.796$. 

One may improve this calculation somewhat by including SU(3) breaking. We note that the 
measured S-wave amplitudes for $\Lambda \to p \pi^-$ and $\Sigma^- \to n \pi^-$ alone 
determine a slightly different value for $x_0$, $x_0=0.835$ having practically no effect on $F$. 
The relation
\beq\label{eqn:rel}
A(\Xi^-_b \to \Lambda_b \pi^-) = -\frac{1}{2\sx}A(\Lambda \to p \pi^-)
- \frac{3}{4}A(\Sigma^- \to n \pi^-)~,
\eeq
and experimental values of the amplitudes on the right-hand side imply 
\beq\label{eqn:Xib}
A(\Xi^-_b \to \Lambda_b \pi^-)= \pm1.75 \pm 0.26~.
\eeq
In the three amplitudes occurring in (\ref{eqn:rel}) an $s$ quark occurs in the 
decaying baryons taking part in the transition but not as a spectator. 
This leads to a common redefinition of $x_0$ which now includes SU(3) breaking.
While the value (\ref{eqn:Xib}) includes this effect of SU(3) breaking we have attributed to it
an uncertainty of $15\%$ caused by assuming octet dominance and by neglecting the effect 
of the heavy $b$ quark on properties of the light diquarks.

The considerations and calculation leading to (\ref{eqn:Xib}) apply also to the contribution 
of the transition $s \to u (\bar u d)$ to the S-wave amplitude for $\Xi^0_c \to \Lambda_c \pi^-$. 
Here one replaces a spectator $b$ quark in $\Xi^-_b$ and $\Lambda_b$ by a $c$ quark in 
$\Xi^0_c = csd$ and $\Lambda_c = cud$, assuming that the $c$ quark mass  is much heavier 
than the light $u, d$ and $s$ quarks.  In this approximation we have
\beq\label{eqn:XicXib}
A_{s \to u\bar ud}(\Xi^0_c \to \Lambda_c \pi^-) = A(\Xi^-_b \to \Lambda_b \pi^-)~.
\eeq

\section{$c s \to c d$ CONTRIBUTION TO $\Xi^0_c \to \Lambda_c \pi^-$}\label{sec:cscd}

The S-wave amplitude for $\Xi^0_c \to \Lambda_c \pi^-$ obtains a second contribution from 
an ``annihilation" subprocess  $c s \to c d$ involving an interaction between the $c$ and $s$ 
quarks in the $\Xi^0_c$. We will now present in some detail a method proposed by 
Voloshin~\cite{Voloshin:2000et,Voloshin:1999pz,Voloshin:1999ax} for calculating this amplitude
in the heavy $c$-quark limit in terms of differences among measured total widths of charmed 
baryons. 

The effective weak Hamiltonian responsible for this Cabibbo-suppressed strangeness-changing transition is given by
\beq\label{eqn:CS}
H_W = -\s G_F \cos\theta_C\sin\theta_C\left[(C_+ + C_-)(\bar c_L\gamma_\mu s_L)
(\bar d_L\gamma_\mu c_L) + (C_+ - C_-)(\bar d_L \gamma_\mu s_L)(\bar c_L \gamma_\mu c_L)\right ]~.
\eeq 
In the following we will use values $C_+ = 0.80$ and $C_- = 1.55$ for Wilson
coefficients calculated in a leading-log approximation at a scale $\mu = m_c=
1.4$ GeV corresponding to $\alpha_s(m_c)/\alpha(m_W) = 2.5$. Applying a soft
pion limit and using PCAC, the amplitude due to $cs \to cd$ is given in our
normalization (\ref{eqn:SP}) [which is related to that of Ref.\
\cite{Voloshin:2000et} by a factor $\xi/(G_F m_\pi^2)$] by
\bea\label{eqn:Acscd1}
& & A_{cs\to cd}(\Xi^0_c \to \Lambda_c \pi^-) =
\nonumber\\
& & 
\frac{\s\,\xi}{f_\pi m^2_\pi}\cos\theta_C\sin\theta_C\langle \Lambda_c|
(C_+\hskip-1mm+\hskip-1mmC_-)
(\bar c_L\gamma_\mu s_L)(\bar u_L \gamma_\mu c_L)\,+\,(C_+\hskip-1mm-\hskip-1mmC_-)
(\bar u_L\gamma_\mu s_L)(\bar c_L\gamma_\mu c_L)|\Xi^0_c\rangle
\nonumber\\
& & = \frac{\xi}{2 \s f_\pi m^2_\pi}\,\cos\theta_C\sin\theta_C\,\left[ 0.75\,x - 
2.35\,y\right]~.
\eea
Here $f_\pi = 0.130$ GeV, $\xi \equiv 2m_{\Xi^0_c}/\sqrt{(m_{\Xi^0_c} + m_{\Lambda_c})^2 - m^2_{\pi^-}} = 1.04$~\cite{Agashe:2014kda}. 

In the above one defines two matrix element $x$ and $y$ (of dimension GeV\,$^3$) in which the 
contribution of the axial-current vanishes for a heavy $c$ quark,  
\bea
x & \equiv & -\langle \Lambda_c|(\bar c\gamma_\mu c)(\bar u \gamma_\mu s)| \Xi^0_c\rangle~,
\nonumber\\
y & \equiv  & -\langle \Lambda_c|(\bar c_i\gamma_\mu c_k)(\bar u_k \gamma_\mu s_i)| 
\Xi^0_c\rangle~, 
\eea
where $i, k$ are color indices. 
Using flavor SU(3) one may write these two terms as differences of diagonal matrix elements of four fermion operators, $\langle {\cal O}\rangle_{\psi - \phi} 
\equiv \langle \psi|{\cal O}|\psi\rangle - \langle \phi|{\cal O}|\phi\rangle$, for charmed baryon states belonging to V-spin and U-spin doublets:
\bea
x & = & \half\langle(\bar c\gamma_\mu c)\left[(\bar u\gamma_\mu u) - (\bar s\gamma_\mu s)
\right]\rangle_{\Xi^0_c-\Lambda_c} =
\half\langle(\bar c\gamma_\mu c)\left[(\bar s\gamma_\mu s) - (\bar d\gamma_\mu d)
\right]\rangle_{\Lambda_c-\Xi^+_c}~,
\\
y & = & \half\langle(\bar c_i\gamma_\mu c_k)\left[(\bar u_k\gamma_\mu u_i) - 
(\bar s_k\gamma_\mu s_i)\right]\rangle_{\Xi^0_c-\Lambda_c} =
\half\langle(\bar c_i\gamma_\mu c_k)\left[(\bar s_k\gamma_\mu s_i) - 
(\bar d_k\gamma_\mu d_i)\right]\rangle_{\Lambda_c-\Xi^+_c}~.\nonumber
\eea

Within a heavy quark expansion the quantities $x$ and $y$ can be used  to describe differences 
of inclusive decay rates among the above three charmed baryons. Adding contributions of hadronic 
and semileptonic Cabibbo-favored and singly Cabibbo-suppressed decays one 
finds in the flavor SU(3) limit~\cite{Voloshin:2000et,Voloshin:1999pz,Voloshin:1999ax}:
\bea
\Gamma(\Xi^0_c) - \Gamma(\Lambda_c) & = & \frac{G^2_Fm^2_c}{4\pi}\left(-x\,[\cos^4\theta_C\,C_+C_-
+ \frac{1}{4}\cos^2\theta\sin^2\theta(6C_+C_- + 5C^2_++5C^2_-)] \right.
\nonumber\\
& & + \left.y\,[3\cos^4\theta_CC_+C_- +\frac{3}{4}\cos^2\theta_C\sin^2\theta_C(6C_+C_- - 3C^2_+ 
+ C^2_-) + 2]\right)~,
\nonumber\\
\Gamma(\Lambda_c) - \Gamma(\Xi^+_c) & = & \frac{G^2_Fm^2_c}{4\pi}
\left(-x\,\frac{1}{4}\cos^4\theta_C(5C^2_+ + 5C^2_- - 2C_+C_-) \right.
\nonumber\\
& & + \left. y\,[\frac{3}{4}\cos^4\theta_C(C^2_- - 3C^2_+ - 2C_+C_-) - 2(\cos^2\theta_C 
- \sin^2\theta_C)] \right)~.
 \eea 
 Substituting the above values of $C_+, C_-$ and $\cos\theta_C=0.97424, 
 \sin\theta_C=0.2253$~\cite{Agashe:2014kda} one has
 \bea
 \Gamma(\Xi^0_c) - \Gamma(\Lambda_c) & = & \frac{G^2_Fm^2_c}{4\pi}[ -1.39\,x + 5.64\,y]~,
 \nonumber\\
 \Gamma(\Lambda_c) - \Gamma(\Xi^+_c) & = & \frac{G^2_Fm^2_c}{4\pi}[ -2.87\,x - 3.15\,y ]~.
 \eea
 
 Eliminating $x$ and $y$ in these equations Eq.\ (\ref{eqn:Acscd1}) now implies
 \beq\label{eqn:Acscd}
A_{cs\to cd}(\Xi^0_c \to \Lambda_c \pi^-) = -\frac{\s\pi\xi\cos\theta_C\sin\theta_C}
{G^2_Fm^2_cm^2_\pi\,f_\pi}\,\left(0.44[\Gamma(\Xi^0_c) - \Gamma(\Lambda_c)]
+ 0.05[ \Gamma(\Lambda_c) - \Gamma(\Xi^+_c)]\right)~.
 \eeq
Using the measured charmed baryon lifetimes~\cite{Agashe:2014kda}
\beq
\tau(\Xi^0_c) = 0.112^{+0.013}_{-0.010}~{\rm ps}~,~~\tau(\Xi_c^+) = 0.442 \pm
0.026~{\rm ps}~,~~ \tau(\Lambda_c) = 0.200 \pm 0.006~{\rm ps}~,
\eeq 
we calculate
\beq\label{eqn:Xicstou}
A_{cs\to cd}(\Xi^0_c \to \Lambda_c \pi^-) = -(1.85 \pm 0.40 \pm 0.40)  \left(\frac{1.4\,{\rm GeV}}{m_c}\right)^2~.
\eeq
The first (symmetrized) error corresponds to errors in lifetime measurements,
while the second one is associated with uncertainties due to SU(3) breaking and
due to a finite $c$-quark mass.  We checked that replacing the Wilson
coefficients $C_\pm$ by values calculated beyond the leading-log approximation,
$C_+ = 0.80, C_- = 1.63$~\cite{Buchalla:1995vs}, has a negligible effect on 
the central value.

\section{DECAY RATES OF $\Xi^0_c\to \Lambda_c \pi^-$ AND $\Xi^+_c\to \Lambda_c \pi^0$} 
\label{sec:total} 

Combining Eqs.~(\ref{eqn:Xib}), (\ref{eqn:XicXib}) and (\ref{eqn:Xicstou}) and
adding errors in quadrature we find for $m_c=1.4$ GeV and destructive
interference
\beq\label{eqn:Ampd}
A(\Xi_c^0 \to \Lambda_c\pi^-) = |A_{s \to u\bar ud}(\Xi_c^0\to \Lambda_c
\pi^-)| + A_{cs\to cd}(\Xi_c^0\to \Lambda_c \pi^-) = - 0.10 \pm 0.62~,
\eeq
while for constructive interference we find
\beq \label{eqn:Ampc}
A(\Xi_c^0\to \Lambda_c \pi^-) = - |A_{s \to u\bar ud}(\Xi_c^0\to \Lambda_c
\pi^-)| + A_{cs\to cd}(\Xi_c^0\to \Lambda_c \pi^-) = - 3.60 \pm 0.62~.
\eeq 
In the former case the small central value of the amplitude is the result of
cancellation between two real contributions of approximately equal magnitudes
but opposite signs. 

In principle each of the two terms in the above two equations could involve a
phase due to final state strong interactions.  A final state interaction one
might anticipate in S-wave $\Xi_c \to \Lambda_c \pi$ or $\Xi_b \to \Lambda_b
\pi$ would be the effect of $\Sigma_c^*$ or $\Sigma_b^*$. However, their parity
is wrong for such contributions. Final state interactions are negligible in 
these heavy baryon decays for the same reason they are small in S-wave
nonleptonic hyperon decays. This is demonstrated by the well-fitted real
amplitudes in Table I and by a triangle relation which follows from more
general considerations~\cite{Lee:1964zzc,Sugawara:1964zz},
\beq
2 A(\Xi^- \to \pi^- \Lambda) + A(\Lambda \to \pi^- p) = -
 (3/2)^{1/2} A(\Sigma^- \to \pi^- n)~,
\eeq
which holds best for real values. The second term in $\Xi^0_c \to \Lambda_c
\pi^-$ due to $cs \to cd$ is real and negative, given in (\ref{eqn:Acscd}) in
terms of width differences among charmed baryons.  
 
For constructive interference, the branching fraction is predicted by 
Eq.\,(\ref{eqn:rate}) to be
${\cal B}(\Xi_c^0 \to \Lambda_c \pi^-) = (1.94 \pm 0.70)\times 10^{-3}$.
This branching ratio is somewhat smaller than that
of the corresponding $\Xi^-_b$ decay, ${\cal B}(\Xi^-_b \to \Lambda_b \pi^-) = 
(6.00 \pm 1.81)\times 10^{-3}$, 
calculated using~(\ref{eqn:Xib}) and the $\Xi^-_b$ lifetime which is roughly an order 
of magnitude larger than $\tau(\Xi^0_c)$~\cite{Agashe:2014kda}. 
For destructive interference, at 90\% c.l. it is less than $\sim 10^{-4}$.
The amplitude for $\Xi_c^+ \to \Lambda_c \pi^0$ is related to that for $\Xi_c^0
\to \Lambda_c \pi^-$ by the $\Delta I = 1/2$ rule, which holds for both
contributions.  Consequently, the partial decay rate is half that for
$\Xi_c^0 \to \Lambda_c \pi^-$. 
Because of the larger lifetime of  the $\Xi_c^+$, which 
is about four times that of $\Xi_c^0$, the corresponding branching fraction
is predicted to be about two times larger, $(3.86 \pm 1.35)\times 10^{-3}$
for constructive interference or less than about $2 \times 10^{-4}$ for destructive
interference.

\section{CONCLUSIONS} \label{sec:con}

We have discussed the heavy-flavor-conserving decays $\Xi_c^0 \to \Lambda_c
\pi^-$ and $\Xi_c^+ \to \Lambda_c \pi^0$ within the context of current
algebra, taking separate account of amplitudes governed by the subprocesses
$s \to u \bar u d$ and $cs \to cd$.  We have used a previous result for
$\Xi_b^- \to \Lambda_b \pi^-$ to obtain the former amplitude, while updating
an estimate by Voloshin for the latter.  The relative signs of the amplitudes
are not determined. For constructive interference, we predict ${\cal B}(\Xi_c^0
\to \Lambda_c \pi^-) = (1.94 \pm 0.70)\times 10^{-3}$ with half the rate
and twice the branching fraction for $\Xi_c^+ \to \Lambda_c \pi^0$.  For
destructive interference, the former branching fraction is expected to be less
than about $10^{-4}$ or twice that for the latter.

\section*{ACKNOWLEDGMENTS}

The work of J.L.R. was supported in part by the United States Department of
Energy through Grant No.\ DE-FG02-13ER41598.

\end{document}